# Raman and Photoluminescence Study of Dielectric and Thermal Effects on Atomically Thin MoS$_2$

Rusen Yan[1,4], Simone Bertolazzi[2], Jacopo Brivio[2], Tian Fang[1], Aniruddha Konar[1], A. Glen Birdwell[3], N. V. Nguyen[4], Andras Kis[2], Debdeep Jena[1], and Huili Grace Xing[1,*]

[1] Department of Electrical Engineering, University of Notre Dame, Notre Dame, IN 46556, USA

[2] Electrical Engineering Institute, Ecole Polytechnique Federale de Lausanne(EPFL), CH-1015 Lausanne, Switzerland

[3] Sensors and Electron Devices Directorate, U.S. Army Research Laboratory, Adelphi, Maryland 20783, USA

[4] Semiconductor and Dimensional Metrology Division, National Institute of Standards and Technology, Gaithersburg, Maryland 20899, USA

[*] Corresponding author e-mail: hxing@nd.edu

ABSTRACT: Atomically thin two-dimensional molybdenum disulfide ($MoS_2$) sheets have attracted much attention due to their potential for future electronic applications. They not only present the best planar electrostatic control in a device, but also lend themselves readily for dielectric engineering. In this work, we experimentally investigated the dielectric effect on the Raman and photoluminescence (PL) spectra of monolayer $MoS_2$ by comparing samples with and without $HfO_2$ on top by atomic layer deposition (ALD). Based on considerations of the thermal, doping, strain and dielectric screening influences, it is found that the red shift in the Raman spectrum largely stems from modulation doping of $MoS_2$ by the ALD $HfO_2$, and the red shift in the PL spectrum is most likely due to strain imparted on $MoS_2$ by $HfO_2$. Our work also suggests that due to the intricate dependence of band structure of monolayer $MoS_2$ on strain, one must be cautious to interpret its Raman and PL spectroscopy.

Recently, successful mechanical exfoliation of material down to one-atom thick has inspired intense research interests on two-dimensional (2D) crystals.[1, 2] Graphene, for example, has been the focus of recent research because of the novel Dirac Fermion particle nature of electrons in graphene and its potential for electronic and optical applications.[3, 4] However, the lack of an intrinsic bandgap substantially limits the graphene applications in electronic transistors. In this context, single layer molybdenum disulfide ($MoS_2$) films, possessing a direct bandgap of 1.8 eV, have become attractive.[5, 6] Single-layer $MoS_2$ based transistors with a high on/off ratio of $10^8$ has been demonstrated recently using $HfO_2$ as a top gate dielectric.[7] Few-layer $MoS_2$ devices with both n-type channel and p-type inversion channel have also been demonstrated.[8] In these atomically thin two dimensional (2D) materials, though the atoms are confined in a plane, the electric field originating from charges in the 2D crystals can leak out to its surroundings. Thus, the dielectric permittivity of the surrounding layers has a profound impact on the electronic and optoelectronic properties of materials with low-dimensionalities. Subsequently, dielectric engineering[9] has been coined to capture this fundamentally novel approach to design functional semiconductor devices, in addition to the well-known band engineering approach in the semiconductor field. Jena *et al.* predicted electron mobility enhancement in 2D and 1D semiconductors encompassed in high-K dielectrics[9] which were also experimentally verified.[7, 10, 11] The dielectric effect has been intensively studied for graphene, in terms of electron transport, Raman spectrum etc.[10, 12, 13] However, there are yet very few reports on high-K dielectric coated $MoS_2$. In this letter, we describe our study on the influence of dielectrics on phonon vibrations of mono- and few- layer $MoS_2$ and photoluminescence (PL) of monolayer $MoS_2$. A red shift was consistently observed in both Raman and PL spectra of $MoS_2$ on sapphire with $HfO_2$ on top in comparison to $MoS_2$ without $HfO_2$ covered. The Raman shift has been attributed to the vibrational Stark effect[14] or the phonon mode softening due to increased carrier concentration, most probably due to positive charges present in $HfO_2$ and near the $HfO_2$/$MoS_2$ interface. The PL shift has been attributed to the strain imparted by the $HfO_2$ on top deposited by atomic layer deposition (ALD). We have also found that the Raman and PL spectra of monolayer $MoS_2$ exhibit a substantial dependence on the excitation laser intensity due to local heating induced thermal expansion of the crystal.[15, 16] This study thus provides an improved understanding of the dielectric effects and thermal properties of the 2D $MoS_2$ crystals, critical for future $MoS_2$ device design and fabrication.

The ultrathin $MoS_2$ films were fabricated from bulk crystals of molybdenite (SPI) by widely used mechanical exfoliation method.[2] Flakes of $MoS_2$ were first deposited onto $SiO_2$/Si wafers coated with polyvinyl (PVA) and polymethyl methacrylate (PMMA). The single and multiple layers of $MoS_2$ films are identified using optical microscope[17] and atomic force microscope (not shown), and then transferred onto target substrates using the method that has been described elsewhere.[18, 19] In Fig. 1 (a) and (b) we show the optical images of two typical $MoS_2$ fakes on sapphire substrates with and without additional 30 nm ALD $HfO_2$ on top, respectively. ALD was performed in a home-built reactor using a reaction of $H_2O$ with tetrakis(dimethylamido)hafnium (Sigma Aldrich) at 200 °C.

The Raman measurements were carried out using a WITec Raman confocal microscope. The Raman spectra presented in this paper were collected using a 488 nm solid-state laser for excitation with the beam focused by a 100x objective lens (the beam diameter is about to be 0.5 – 1 μm [ref]). The characteristic Raman spectra of the monolayer $MoS_2$ flakes highlighted in Fig. 1 (a) and (b) are presented in Fig. 2 (a) with a relatively low laser excitation power of 0.25 mW. Two prominent peaks are observed around 400 $cm^{-1}$ in both samples, corresponding to an in-plane vibration ($E_{2g}^1$) of Mo and S atoms and the out-of-plane vibration ($A_{1g}$) of S atom as shown in the inset. The peak positions are determined by Lorentzian fitting of the peaks. It is observed that the top high-K $HfO_2$ gives rise to an appreciable red shift of ~2.5 $cm^{-1}$ for the $A_{1g}$ mode but has a negligible influence to the $E_{2g}^1$ mode. Also, notice the full width at half maximum (FWHM) for $A_{1g}$ mode is broadened after the deposition of $HfO_2$, indicating the strong modification on phonon vibrations induced by external effects. This behavior is consistently observed in all monolayer $MoS_2$ flakes in contact with $HfO_2$ that we studied. The layer dependent Raman spectra taken on $MoS_2$ with $HfO_2$ on top were also measured and are shown in Fig. 3(a). Plotted in Fig. 3(b) are the frequency differences between the two Raman modes for monolayer and bulk $MoS_2$ with and without $HfO_2$ as well as the layer dependence reported in the literature.[20] It is seen that the red shift induced by $HfO_2$ is most prominent in monolayer $MoS_2$, but weakens with increasing layer thicknesses and disappears in bulk $MoS_2$.

A detailed attribution of the red shift in Raman to the vibrational Stark effect will be presented shortly. First, let us scrutinize the effects of dielectric screening and sample heating by the excitation laser. It has been commonly observed in $MoS_2$ that the $E_{2g}^1$ mode red shifts and the $A_{1g}$ mode blue shifts with increasing layer thicknesses, which has been explained by several mechanisms including dielectric screening.[21, 22] This layer dependent Raman behavior is indeed consistent with our observation shown in Fig.3. However, for monolayer $MoS_2$ with $HfO_2$ on top, a red shift in both Raman modes was observed compared to monolayer $MoS_2$ without $HfO_2$. Therefore, dielectric screening by $HfO_2$ alone most likely cannot explain the Raman shift in monolayer $MoS_2$. It is also well known that higher optical excitation power used in Raman measurements leads to sample local heating thus thermal expansion of the sample lattice, as a result, softening of phonon frequencies.[23, 24] As shown in Fig. 2 (b), this trend is also maintained in our $MoS_2$ samples: both of the two notable peaks soften as varying laser excitation power. Note that as laser power becomes larger than 1mW, the softening of both peaks saturate. In Fig. 2 (c), we showed the laser power dependent Raman peak positions under excitation powers lower than 0.5 mW, where the linear fittings can well characterize the peak position changes, allowing us to extract corresponding zero-power peak positions at room temperature (shown in Table. 1). Note that different types of markers in the figure represent the peak positions extracted from different flakes, showing the reproducibility of our observations. The difference in the slope for the out-of-plane and in-plane modes is possibly due to the different thermal expansion coefficients of $MoS_2$ in the two directions since, intuitively, the 2D crystals can expand more readily in the out-of-plane direction than in plane.[16] The difference in the slopes for the out-of-plane and in-plane modes can be attributed to two reasons: 1) larger thermal expansion coefficient of $MoS_2$ in in-plane than out-of-plane direction; 2) different strains induced by the different thermal expansion coefficients of $MoS_2$, sapphire and $HfO_2$. We estimated that the absorption of laser power by monolayer $MoS_2$ is about 9% while that by sapphire and HfO2 is almost zero considering their larger band gap than the incident photon energy.

Consequently, the local temperature on MoS$_2$ is much higher than those on sapphire and HfO$_2$, which leads to a stronger thermal expansion of MoS$_2$ than those of sapphire and HfO$_2$ as increasing the laser power. This discrepancy of thermal expansion induces the evolution of the strain with the laser power, possibly resulting in the less softening of E$_{2g}^1$ mode considering the fact that E$_{2g}^1$ is much more sensitive than A$_{1g}$ mode to the strain variation[25]. Therefore, we conclude that the red shift in A$_{1g}$ for MoS$_2$ in contact with HfO$_2$ is not a result of thermal effects, because a lower temperature rise is expected in MoS$_2$ with HfO$_2$ on top than the sample without, considering HfO$_2$ can act as a heat dissipation channel. To explain the observed red shift, we invoke a simple harmonic oscillator model of atomic vibration assuming a sheet of positive charge situated at an equivalent distance of $d_0$ in HfO$_2$ above the top S atom plane of MoS$_2$, as sketched in Fig. 4(a). The positive fixed charges present in HfO$_2$ close to the interface can be potentially induced by charge transfer due to formation of Hf-S bond, or oxygen vacancy[26-28] and impurities[29] formed during the ALD deposition. As a consequence, an additional Coulomb potential perpendicular to the MoS$_2$ plane ($x$-direction) arises from the attraction between the negatively charged S atoms and the positive fixed charges in HfO$_2$. Due to the geometrical considerations, we note that this additional potential affects mostly the out-of-plane vibration (A$_{1g}$) while has a minimal effect on the in-plane vibrations (E$_{2g}^1$), which is consistent with our observations. Given that A$_{1g}$ mode involves only S atoms, the restoring constant $k$ corresponding to this mode can be simply described by $k = d^2V(x)/dx^2$ where $V(x)$ is the potential field experienced by the S atoms and $x$ is the atomic displacement from equilibrium. Assuming the effective negative sheet charge near the S-atom plane and the positive sheet charge near the Hf-atom plane are $Z_1 e$ and $Z_2 e$, respectively, the additional Coulomb potential can be written as $\Delta V(x) = -Z_1 Z_2 e^2/4\pi\varepsilon_0 x$, where $\varepsilon_0$ is free space permittivity. Both the parabolic potential arising from the S-Mo-S atomic bond and the Coulomb potential induced by HfO$_2$ are sketched in Fig. 4(b). Due to the presence of Coulomb attraction, the effective spring constant of the harmonic oscillator (curvature of total potential) decreases to be $k' = k + \Delta k$ and $\Delta k = d^2 \Delta V(x)/dx^2 = -2Z_1 Z_2 e^2/4\pi\varepsilon_0 x^3 < 0$. This change of spring constant has been termed as the atomic vibrational stark effect.[30] The appreciable Raman shift observed for MoS$_2$ coated with HfO$_2$ in this study is in contrast to the negligible shift reported for MoS$_2$ on sapphire or on SiO$_2$.[20] We speculate it is because the fixed charge density in HfO$_2$ deposited on top of MoS$_2$ is significantly higher than that in the supporting SiO$_2$ substrate,[13] which is also manifested by a poorer adhesion of MoS$_2$ on SiO$_2$.[20,31] Furthermore, in few-layer MoS$_2$ electrons spread out over all the layers, consequently weakening the Coulomb interaction and its influence on the Raman modes, consistent with the observation shown in Fig. 3(b).

The observed red shift in the Raman spectra in this work is similar to a recent study on electron-phonon coupling in MoS$_2$ by Chakraborty et al.[22]. There, Raman measurements were performed in a monolayer MoS$_2$ gated by a polymer electrolyte and, with increasing electron concentration, an appreciable red shift was observed in $A_{1g}$ but a negligible red shift in $E_{2g}^1$, which was attributed to the electron-phonon coupling in MoS$_2$ supported by a density functional theory (DFT) modeling effort. Based on Chakraborty's results, we estimate the electron concentration increase in HfO$_2$ coated MoS$_2$ to be $6.5 \times 10^{12}$ cm$^{-2}$ assuming a linear slope of $2.6 \times 10^{12}$ cm$^{-2}$/cm$^{-1}$ for the $A_{1g}$ shift with electron concentration. Furthermore, their measurement also show an increase of FWHM from about 5 cm$^{-1}$ to 9 cm$^{-1}$ due to the strengthening of electron-phonon coupling, in excellent agreement with our observation as shown in Fig. 2 (d). This result is largely consistent with our aforementioned simple model: the ALD HfO$_2$ modulation dopes MoS$_2$ with more electrons. The only difference is that in our model the phonon softening arises from an extrinsic out-of-plane dipole interaction, and in Chakraborty's model the electron-phonon coupling is intrinsic to monolayer MoS$_2$. It is worth noting that, since in Chakraborty's experiment electrons electrostatically induced in MoS$_2$ are subject to a strong interaction with the positive charges situated within about 1 nm in the polymer electrolyte gate, one cannot safely exclude the effect of this extrinsic out-of-plane dipole. A future experiment to isolate the effect of electron-phonon coupling can be potentially carried out in chemically doped monolayer MoS$_2$ so that the electron concentration can be varied while keeping the net out-of-plane dipole being zero.

After discussing the impact of dielectric screening, heating and doping on the Raman spectrum of MoS$_2$, we turn our attention to strain. A recent DFT calculation[32] suggested that both the $E_{2g}^1$ and $A_{1g}$ modes red (blue) shift when monolayer $MoS_2$ is under tensile (compressive) strain and the shift of $E_{2g}^1$ is much greater than that of $A_{1g}$; furthermore, the energy bandgap decreases under either tensile or compressive strain. Another DFT study[33] suggested that the bandgap of monolayer $MoS_2$ decreases under tensile strain but increase slightly under compressive strain, and further suggested that ALD $HfO_2$ on top of $MoS_2$ typically imparts tensile stress to $MoS_2$, therefore explaining the experimentally observed decrease in bandgap inferred from the photoluminescence measurement. Beyond that, an experimental effort applying uniaxial tensile strain by H. J. Conley et al reveals that the bandgap of monolayer MoS2 linearly decreases as the tensile strain increases with a linear coefficient of 45meV/%.[34] Though those studies are not totally consistent, all point out that under tensile strain both the Raman modes and PL peak should red shift. To this end, we carried out the PL measurements on monolayer $MoS_2$ with and without $HfO_2$ using a continuous-wave excitation at 633 nm. The typical PL spectra are shown in Fig. 5(a). A red shift of ~30 meV is indeed observed for the monolayer $MoS_2$ with ALD $HfO_2$ on top, which indicates a tensile strain of less than 0.67%.[34] The excitation power dependence of the PL peak for both samples is summarized in Fig. 5(b), showing the PL peaks red shift linearly with the increasing power but the slope for the $HfO_2$-covered $MoS_2$ is about 5x smaller. This difference can be attributed to the thermal expansions difference between monolayer $MoS_2$ and $HfO_2$: the oxide on top can hinder expansion of $MoS_2$. It is also possible that the temperature rise of $MoS_2$ with $HfO_2$ is smaller due to cooling via $HfO_2$, but the $MoS_2$ temperature needs to be accurately determined to understand the contribution of this effect. On the other hand, the dielectric environment is known to impact PL. Keldysh predicted in 1979[35] and subsequently verified by experiments[36]: high-K dielectrics surrounding nanoscale thin semiconductor films reduce the Coulombic interaction between electrons and holes thus reducing the exciton binding energy. If this screening effect dominates the PL, a blue shift is expected since the PL peak energy can be estimated by subtracting the exciton binding energy from the bandgap, which is again contrary to our observation. Therefore, the red shift in PL most probably arises from strain imparted on monolayer $MoS_2$ by ALD $HfO_2$. Next we scrutinize whether strain is also the dominating factor in the Raman spectra shift. The DFT study[32] suggested that strain induces a larger shift in the in-plane mode $E_{2g}^1$, but we observe the opposite: a much large shift in the out-of-plane mode $A_{1g}$. Our experimental observation directly implies that $HfO_2$ introduces a much higher force constant change for the out-of-plane mode than the in-plane mode, more consistent with our proposed model and its geometric characteristics. Based on all the above considerations on heating, doping, strain and dielectric screening, we suggest that the red shift in Raman largely stems from modulation doping of $MoS_2$ by ALD $HfO_2$.

In summary, we have compared Raman and PL spectroscopy of monolayer $MoS_2$ with and without ALD $HfO_2$ on top to understand the dielectric and thermal effects on two-dimensional crystals. It is found that dielectric screening is not the dominating factor in the $HfO_2$ induced shift observed in Raman or PL. Instead, modulation doping and strain induced by $HfO_2$ are most likely responsible for the shift in Raman and PL, respectively. Our study suggests that the dielectric environment has a profound influence on the properties of ultrathin 2D crystals, and that the dominant factor needs to be very carefully isolated since multiple mechanisms can be present. We believe that the work presented in this letter could be extended to other two-dimensional materials and enrich the knowledge of these promising materials.

ACKNOWLEDGMENT The authors acknowledge the support from NSF (CAREER ECCS-084910, ECCS-1232191, monitored by Anupama Kaul), AFOSR (FA9550-12-1-0257, monitored by James Hwang), the Midwest Institute of Nanoelectronics Discovery (MIND), the Center for Nanoscience and Technology (NDnano) at the University of Notre Dame, Swiss National Science Foundation (Grants 132102 and 138237) and the Swiss Nanoscience Institute (NCCR Nanoscience).

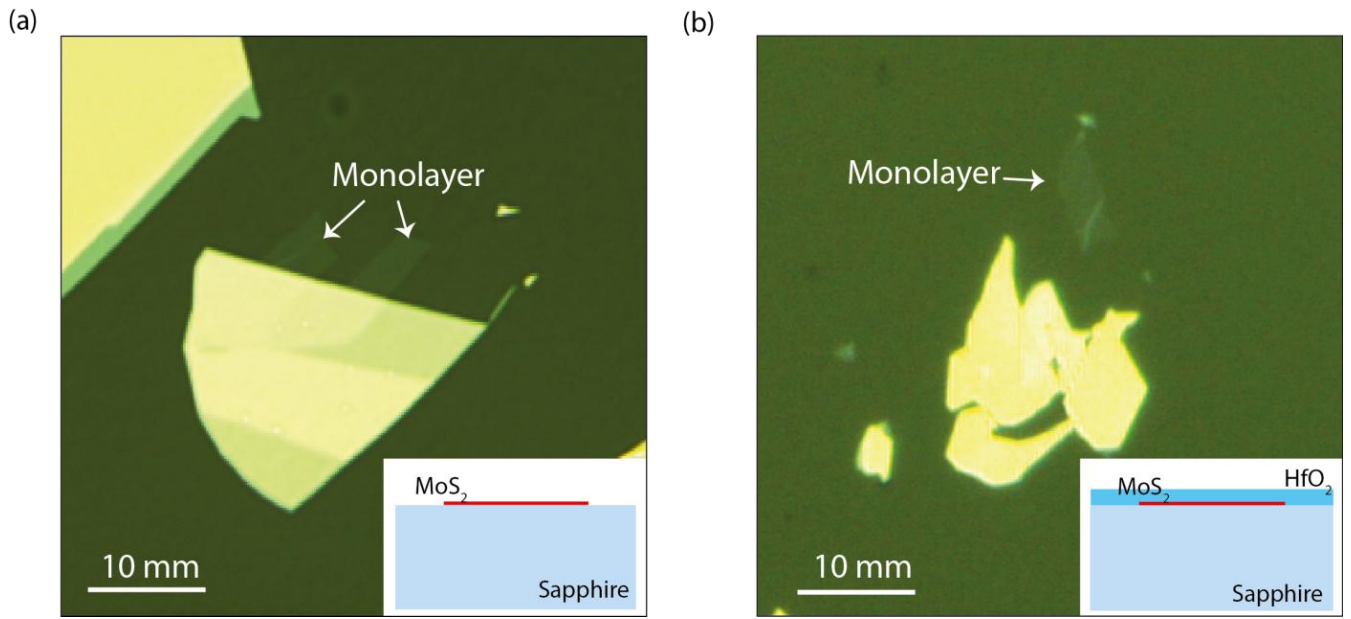

**Figure 1.** Optical images of exfoliated MoS$_2$ flakes placed on sapphire substrate with (a) and without (b) HfO$_2$ on top. Insets show corresponding sample cross-sections.

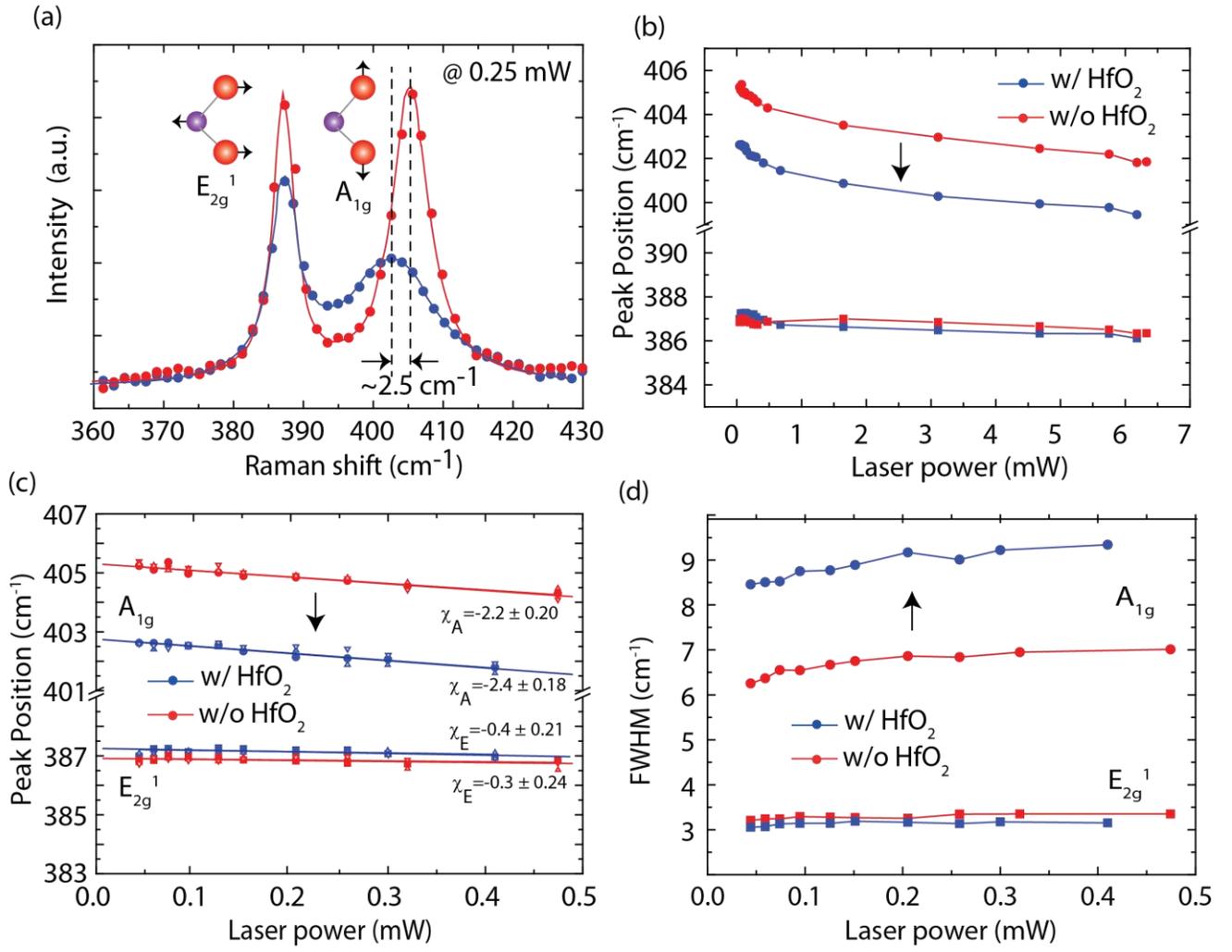

**Figure 2.** (a) Characteristic Raman spectra of the monolayer $MoS_2$ flakes shown in Fig. 1(a) and (b) at a laser excitation power of 0.25 mW. (b) and (c) Raman peak positions of $A_{1g}$ and $E_{2g}^1$ modes at different excitation powers. (c) shows the Raman peak positions at low excitation power (<0.5mW). Different types of markers represent peak positions extracted on different flakes. $\chi_A$ and $\chi_E$ are respectively the slope of linear fitting for $A_{1g}$ and $E_{2g}^1$ peaks ($cm^{-1}/mW$). (d) FWHM of $A_{1g}$ and $E_{2g}^1$ modes at low excitation powers. Note that in all plots, red and blue markers respectively represent monolayer $MoS_2$ flakes without and with $HfO_2$ covered on top.

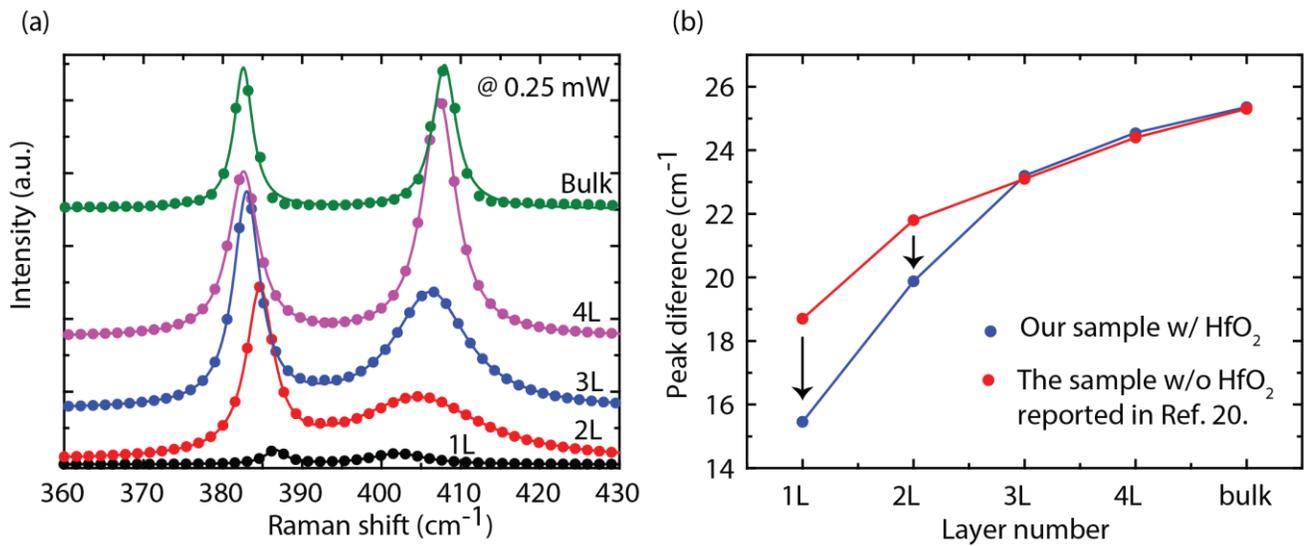

**Figure 3.** (*a*) Raman spectra of $MoS_2$ with varying layer thicknesses on sapphire with $HfO_2$ on top. (*b*) Raman frequency difference between $A_{1g}$ and $E_{2g}^1$ as a function of layer number for $HfO_2$ coated $MoS_2$. Also shown are the reported values in the literature[20] and the measured Raman frequency difference on monolayer and bulk $MoS_2$ without $HfO_2$ in this study.

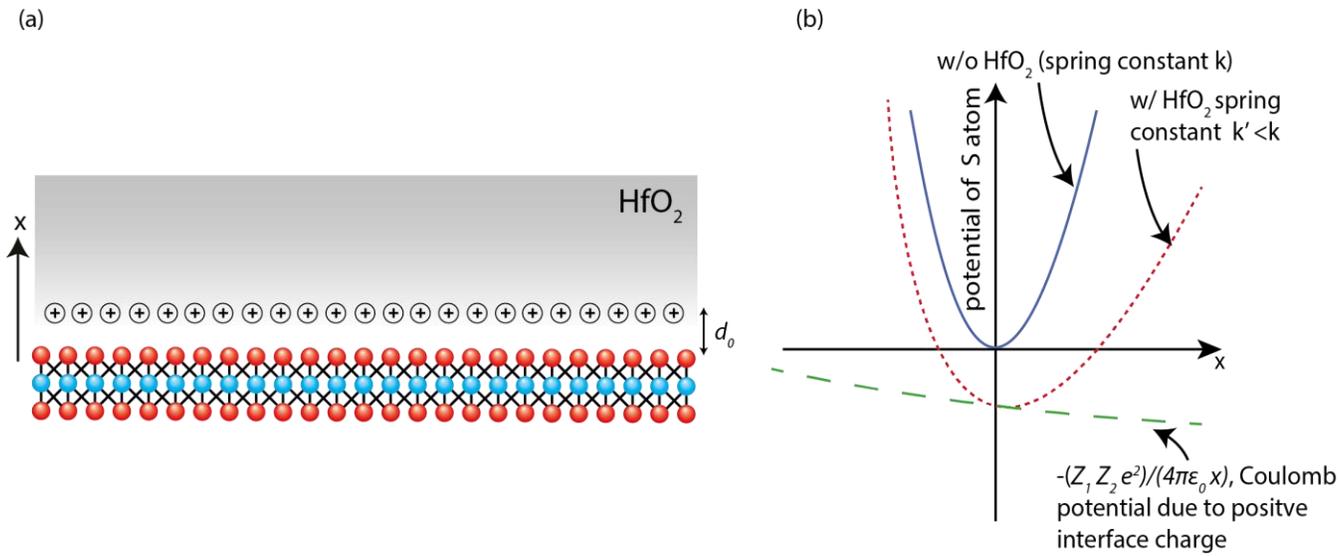

**Figure 4.** (*a*) Schematic of the $HfO_2$/$MoS_2$ structure and the positive charges in $HfO_2$ are assumed to be located at $d_0$ away from the top S atom plane. (*b*) Potential configuration of the top S atom. The Coulomb potential exerted by the positive charges near the $HfO_2$/$MoS_2$ interface alters the original potential, weakens the restoration constant thus softening the phonon frequency.

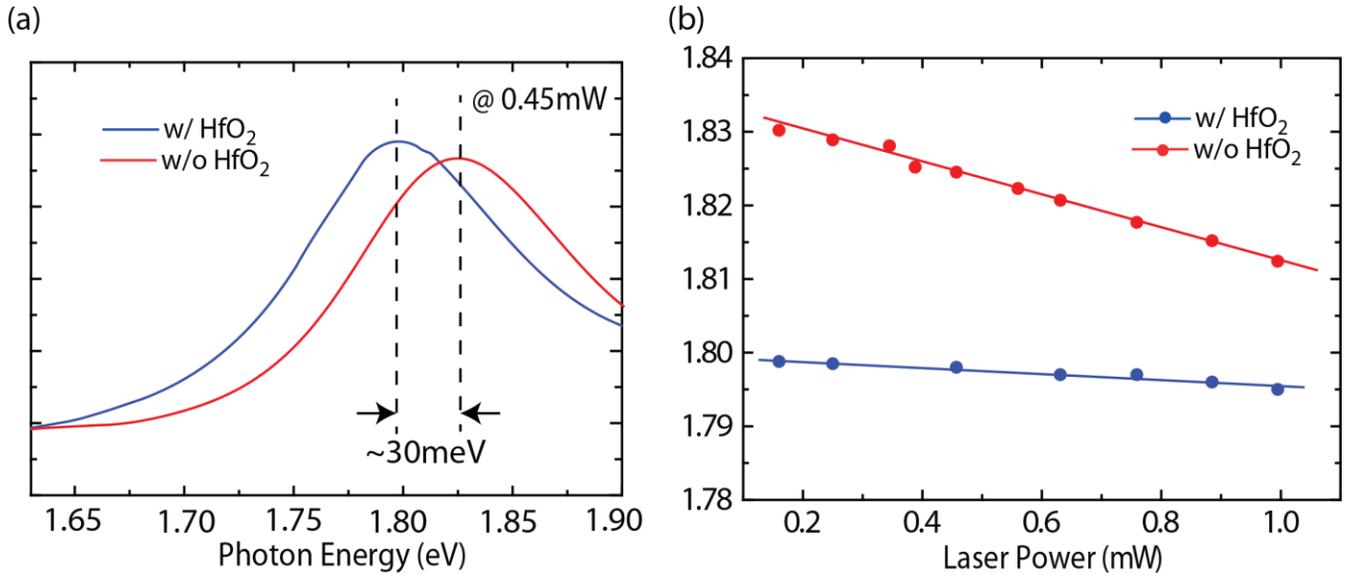

**Figure 5.** (a) PL spectra of the monolayer MoS$_2$ with and without HfO$_2$ on top. (b) PL peak position for the monolayer MoS$_2$ under various excitation powers. η represents the slope of the linear fit of the PL peak position as a function of the excitation laser power.

Table. 1  Extracted peak positions of $E_{2g}^1$ and $A_{1g}$ modes at zero laser power

|  | $E_{2g}^1$ (cm$^{-1}$) | $A_{1g}$ (cm$^{-1}$) |
|---|---|---|
| MoS$_2$ w/o HfO$_2$ | 386.9 | 405.3 |
| MoS$_2$ w/ HfO$_2$ | 387.2 | 402.8 |